\documentclass[conference]{IEEEtran}
\IEEEoverridecommandlockouts
\usepackage{cite}
\usepackage{amsmath,graphicx,url,times,booktabs,tabularx,amsfonts,siunitx}
\usepackage{algorithmic}
\usepackage{graphicx}
\usepackage{textcomp}
\usepackage{xcolor}
\usepackage{url}
\usepackage{multirow}
\usepackage{siunitx}
\usepackage{microtype}
\usepackage{soul}
\usepackage{color}
\usepackage[colorlinks,linkcolor=blue,]{hyperref}

\def\BibTeX{{\rm B\kern-.05em{\sc i\kern-.025em b}\kern-.08em
    T\kern-.1667em\lower.7ex\hbox{E}\kern-.125emX}}
\begin{document}

\let\OLDthebibliography\thebibliography
\renewcommand\thebibliography[1]{
  \OLDthebibliography{#1}
  \setlength{\parskip}{0pt}
  \setlength{\itemsep}{2.4pt plus 0.1ex}
}

\title{Leveraging Pre-trained BERT for Audio Captioning}

\author{\IEEEauthorblockN{1\textsuperscript{st} Given Name Surname}
\IEEEauthorblockA{\textit{dept. name of organization (of Aff.)} \\
\textit{name of organization (of Aff.)}\\
City, Country \\}}

\author{
      \IEEEauthorblockN{
      Xubo Liu$^{1}$,
      Xinhao Mei$^{1}$,
      Qiushi Huang$^{2}$,
      Jianyuan Sun$^{1}$,
      Jinzheng Zhao$^{1}$,
      Haohe Liu$^{1}$,  \\
      Mark D. Plumbley$^{1}$,
      Volkan Kılıç$^{3}$,
      Wenwu Wang$^{1}$
     }
     \\
      \IEEEauthorblockN{$^1$Centre for Vision, Speech and Signal Processing (CVSSP), University of Surrey, UK}
      \IEEEauthorblockN{$^2$Department of Computer Science, University of Surrey, UK}
      \IEEEauthorblockN{$^3$Department of Electrical and Electronics Engineering, Izmir Katip Celebi University, Turkey}
      }
\maketitle

\begin{abstract}
Audio captioning aims at using language to describe the content of an audio clip. Existing audio captioning systems are generally based on an encoder-decoder architecture, in which acoustic information is extracted by an audio encoder and then a language decoder is used to generate the captions. Training an audio captioning system often encounters the problem of data scarcity. Transferring knowledge from pre-trained audio models such as Pre-trained Audio Neural Networks (PANNs) have recently emerged as a useful method to mitigate this issue. However, there is less attention on exploiting pre-trained language models for the decoder, compared with the encoder. BERT is a pre-trained language model that has been extensively used in natural language processing tasks. Nevertheless, the potential of using BERT as the language decoder for audio captioning has not been investigated. In this study, we demonstrate the efficacy of the pre-trained BERT model for audio captioning. Specifically, we apply PANNs as the encoder and initialize the decoder from the publicly available pre-trained BERT models. We conduct an empirical study on the use of these BERT models for the decoder in the audio captioning model. Our models achieve competitive results with the existing audio captioning methods on the AudioCaps dataset. 
\end{abstract}


\section{Introduction}
Audio captioning is the task of generating a text description for an audio clip, which has various potential applications. For example, audio captioning can be used to generate text descriptions of sounds to help the hearing impaired in understanding an acoustic environment. Audio captioning has attracted increasing interest in the fields of acoustic signal processing and natural language processing (NLP).

Existing audio captioning systems are mostly based on an encoder-decoder architecture \cite{drossos2017automated, koizumi2020transformer, mei2021audio, mei2021encoder, liu2021cl4ac, yuan2021_t6, mei2021diverse}, in which acoustic information is extracted by an audio encoder, and then a language decoder is used to generate text descriptions. Training of an audio captioning system often encounters the problem of data scarcity. AudioCaps \cite{kim2019audiocaps} is the largest public dataset for audio captioning research, however, it only has about 50k audio clips with one reference caption. Compared with the popular image captioning datasets such as MS COCO ($\sim$123k images) \cite{mscoco} and Conceptual Captions ($\sim$3.3M images) \cite{conceptualcaptions}, the scale of the existing audio captioning dataset is much smaller. This can limit the performance of an audio captioning model in generating consistent natural language description.

To address the data scarcity issue of audio captioning, transferring knowledge from pre-trained audio models has been widely investigated. Xu et al. \cite{xu2021investigating} propose an approach that uses transfer learning to exploit local and global information from audio tagging and acoustic scene classification, respectively. Pre-trained Audio Neural Networks (PANNs) \cite{kong2020panns} are the models pre-trained on AudioSet \cite{gemmeke2017audio}, which have achieved great success as the encoder \cite{xinhao2021_t6, liu2021cl4ac, yuan2021_t6, mei2021diverse} in the audio captioning system. Nevertheless, compared with the audio encoder, there is less attention on exploiting pre-trained NLP models for the language decoder in the audio captioning model. 

Koizumi et al. \cite{koizumi2020audio} used a frozen Generative Pre-Training model (GPT-2) \cite{radford2019language} with the retrieval of similar captions in the dataset. This method generates accurate results using ground-truth similar captions, whereas using the retrieved captions leads to degraded performance. Gontier et al. \cite{gontierautomated} proposed an approach for audio captioning by fine-tuning Bidirectional and Auto-Regressive Transformers (BART) \cite{lewis2019bart} with AudioSet \cite{gemmeke2017audio} tags as text conditions in the BART encoder, which achieved the state-of-the-art result on AudioCaps \cite{kim2019audiocaps}. However, the performance of this method is highly dependent on the audio tagging model. In addition, the adaptation of BART (12 layers in both the encoder and decoder) results in a large number of training parameters ($\sim$400 million). 

BERT \cite{devlin2018bert}, which stands for Bidirectional Encoder Representations from Transformers, is an NLP model pre-trained on large-scale text datasets, which has been extensively used as strong baselines on many natural language understanding (NLU) benchmarks \cite{williams2017broad, rajpurkar2018know}. Recently, BERT has been exploited as the decoder in sequence-to-sequence models and has achieved state-of-the-art results on several Natural Language Generation (NLG) tasks such as Machine Translation and Text Summarization \cite{rothe2020leveraging}. Weck et al. \cite{weck2021evaluating} used BERT embeddings for the decoder in the audio captioning model. However, using only word embedding layers may not fully utilize the linguistic knowledge of the pre-trained BERT model. In summary, the potential of using BERT for audio captioning has not been well studied in the literature.


In this paper, we investigate the exploitation of pre-trained BERT models for the decoder in the audio captioning model. We propose an encoder-decoder model in which PANNs are used as the audio encoder, and the pre-trained BERT is used in the decoder. To bridge the language decoder and the audio encoder, we add cross-attention layers with randomly initialized weights in the decoder, but retain the pre-trained weights from BERT models for other layers in the decoder. In this way, the knowledge gained from the pre-trained BERT model can be transferred to the audio captioning decoder. We conduct an empirical study for the utility of various pre-trained BERT model such as BERT \cite{devlin2018bert}, Compact BERT \cite{turc2019well} and RoBERTa \cite{liu2019roberta} on the AudioCaps dataset. The experimental results demonstrate the efficacy of the pre-trained BERT models for audio captioning. Our proposed models achieve competitive results, as compared with existing audio captioning methods. 

The remainder of this paper is organized as follows. The next section introduces our proposed method. Section \ref{sec:Experiments} presents experimental setup. Section \ref{sec:Results} shows experimental results on the AudioCaps datasets. Conclusions are given in Section \ref{sec:conclusion}.



\section{Proposed Method}
\label{sec:methods}

\begin{figure}[!t]
    \centering
    \includegraphics[width=\columnwidth]{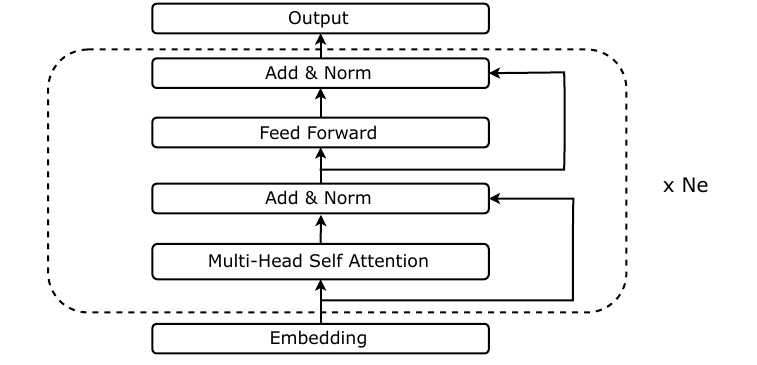}
    \caption{The structure of BERT, where $N_e$ is the number of encoder blocks.}
    \label{fig:network1}
\end{figure}

\begin{figure}
    \centering
    \includegraphics[width=\columnwidth]{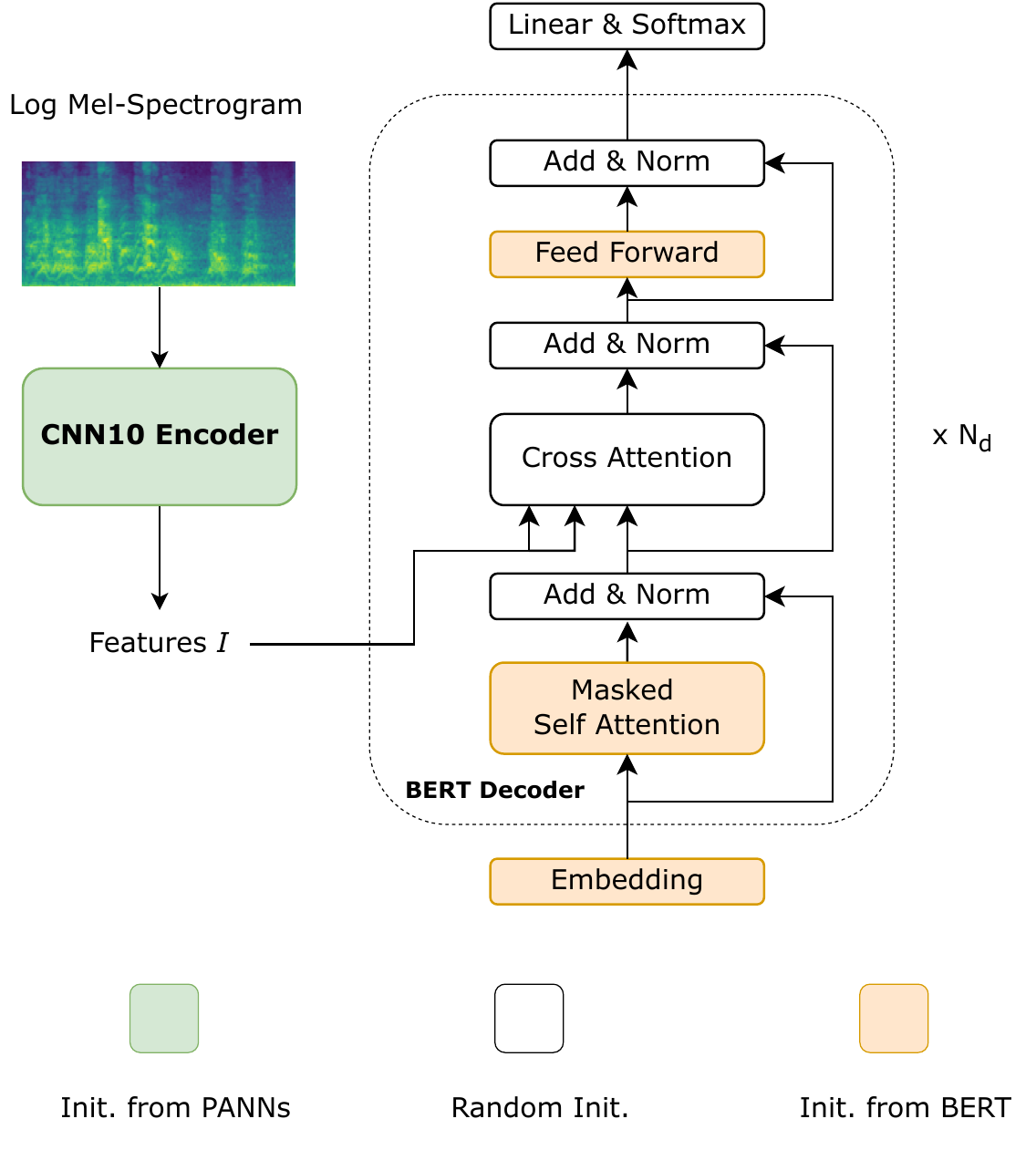}
    \caption{The architecture of the proposed model using PANNs (CNN10) encoder and BERT decoder. The green, white, and orange blocks represent that the weights are initialized with parameters learned from PANNs, randomly initialized, and initialized from BERT, respectively. Here, $N_d$ is the number of BERT decoder blocks.}
    \label{fig:network2}
\end{figure}

Our proposed audio captioning model is composed of PANNs based encoder and BERT based decoder. In this section, we first introduce the pre-trained BERT model, as depicted in Fig. \ref{fig:network1}. Then, we describe the audio encoder of our model, PANNs. Lastly, we discuss the BERT based language decoder in the audio captioning model. Fig. \ref{fig:network2} visualizes the overall architecture of our proposed model.

\subsection{Pre-trained BERT models}

BERT \cite{devlin2018bert} is based on a number of Transformer encoder blocks, where each block contains a multi-head bidirectional self-attention layer followed by a feed-forward layer. Each encoder block is equipped with residual connections and layer normalization. BERT is pre-trained on BooksCorpus \cite{zhu2015aligning} and English Wikipedia using Masked Language Modeling (MLM) and Next Sentence Prediction (NSP) tasks. MLM aims to predict the masked tokens in the input sentence, while NSP aims to predict whether the input two sentences are paired. Pre-training on large datasets using these two tasks offers BERT the capabilities to capture linguistic information such as syntactic and semantic content.

In this work, we investigate three types of publicly available pre-trained BERT models: BERT \cite{devlin2018bert}, Compact BERT \cite{turc2019well}, and RoBERTa \cite{liu2019roberta}. Compact BERT is a compressed version of BERT by knowledge distillation, with a smaller architecture. RoBERTa is built on BERT and uses different pre-training strategies, which shows better performance than BERT. The details of these BERT models are described in Table \ref{tab:bert check}.

\begin{table}[]
\centering
\caption{Configurations of BERT models used in this work.}
\label{tab:bert check}
\begin{tabular}{|c|c|c|c|c|}
\hline
Type                          & Model   & Layer & Head & Hidden \\ \hline
\multirow{3}{*}{Compact BERT} & BERT\_tiny    & 2     & 2    & 128      \\ \cline{2-5} 
                              & BERT\_mini    & 4     & 4    & 256      \\ \cline{2-5}
                              & BERT\_medium  & 6     & 8    & 512      \\ \hline
BERT                 & BERT\_base    & 12    & 12   & 768      \\ \hline
RoBERTa                       & RoBERTa\_base  & 12     & 12    & 768     \\ \hline
\end{tabular}
\end{table}

\subsection{PANNs encoder}
PANNs \cite{kong2020panns} demonstrated powerful capabilities in extracting features of audio signals for audio recognition tasks such as audio tagging. In this work, we use the CNN10 model in PANNs as the audio encoder. The CNN10 consists of four convolutional blocks, each with two convolutional layers with a kernel size of $3 \times 3$. Batch normalization and ReLU are used after each convolutional layer. The number of channels per convolutional block is \num{64}, \num{128}, \num{256} and \num{512}. An average pooling layer with kernel size $2 \times 2$ is applied for down-sampling. Global average pooling is applied along the frequency axis after the last convolutional block, followed by two fully-connected layers to align the dimension of the output with the hidden dimension $D$ of the decoder. The CNN10 encoder takes the log mel-spectrogram of an audio clip as the input and outputs the features $I \in \mathbb R^{T \times D}$, where $T$ and $D$ represent the number of time frames and the dimension of the spectral feature at each time frame, respectively.

\subsection{BERT decoder}
To use BERT as the language decoder in the audio captioning model, we make two adjustments. First, we modify the bidirectional self-attention used in the original BERT model, which considers both past and future context, to unidirectional self-attention by exploiting only the past contexts. This is because the bidirectional structure does not fit the language decoder. Second, the cross-attention layers are added after the self-attention layers to bridge the audio encoder and the language decoder. The cross-attention layer has two inputs, the encoder output $I \in \mathbb R^{T \times D}$ and the current state of the decoder $H \in \mathbb R^{N \times D}$, where $N$ is the number of tokens already decoded, and $D$ is the number of decoder hidden dimension. The cross-attention is calculated as:
\begin{equation}
  {\rm CrossAttn}(H, I) = {\rm Attn}( H, I, I),
\end{equation}
\begin{equation}
  \label{eqn:attention1}
  {\rm Attn}(Q,K,V) = {\rm Softmax}\left (\frac{(W^qQ)(W^kK)^T} {\sqrt{d}}\right)W^vV,
\end{equation}
where $W^q,W^k, W^v$ are three learnable matrices and $d$ is a scaling factor. After that, we apply the add \& norm operation, which contains a residual connection and layer normalization and can be written as:
\begin{equation}
  \label{eqn:attention2}
  {\rm LayerNorm}({\rm CrossAttn}(H, I) + H).
\end{equation}
The cross-attention layers are added with randomly initialized weights, while the other layers retain the pre-trained weights from BERT to transfer the NLP knowledge from BERT.

\section{Experiments}
\label{sec:Experiments}

\subsection{Dataset}
AudioCaps \cite{kim2019audiocaps} is the largest public audio captioning dataset with around 50k audio clips sourced from AudioSet \cite{kim2019audiocaps}. AudioCaps is divided into three splits: training, validation and test. Each audio clip in the training set contains one human-annotated caption, while each clip in the validation and test set has five captions. Since some audio clips are now missing from YouTube, all our experiments are conducted on the version we downloaded, which contains \num{49274} audio clips in the training set, \num{494} clips in the validation set, \num{957} clips in the test set.

\subsection{Audio processing}
We use the original sampling rate of \SI{32000}{\Hz} to load audio data and the mel-spectrogram as the input to our model. Specifically, a \num{64}-dimensional log mel-spectrogram is calculated using the short-time Fourier transform with a Hanning window of \num{1024} samples, and a hop size of \num{512} samples. SpecAugment \cite{park2019specaugment} is used for data augmentation.

\subsection{Text processing}
We converted all captions in the AudioCaps dataset to lower case and removed punctuation. Two special tokens ``\texttt{\textless soc\textgreater}'' and ``\texttt{\textless eoc\textgreater}'' are added to the start and end of each caption. We tokenize our text corpus using the WordPiece \cite{wu2016google} to match the BERT pre-trained vocabulary ($\sim$30k tokens).

\subsection{Training procedure}
We trained the proposed model using Adam \cite{kingma2014adam} optimizer with a batch size of \num{32}. Warm-up is used in the first \num{5} epochs to increase the learning rate to the initial learning rate. The learning rate is then decreased to \num{1}/\num{10} of itself every \num{10} epochs. Dropout with a rate of \num{0.2} is applied in the BERT decoder to mitigate the over-fitting problem. To stabilize the training, we share the weights between the input embedding layer and the output token classification layer in the BERT decoder. We train the model for 30 epochs on the AudioCaps training set, with an initial learning rate of \num{5e-5} for BERT\_base and RoBERTa\_base (as introduced in Table \ref{tab:bert check}) and \num{5e-4} for other BERT configurations. Validation is carried out after every training epoch, and we save the model with the best performance on the validation set. For each experiment, we repeat three times and report their average performance.

\subsection{Evaluation}
During the inference stage, the mel-spectrogram along with the token ``\texttt{\textless soc\textgreater}'' are fed into the encoder and decoder separately to generate the first token. Then, the following tokens are predicted based on the previously generated tokens until the token ``\texttt{\textless eoc\textgreater}'' or the maximum length (50 tokens in our experiments) is reached. The beam search strategy \cite{tillmann2003word} with a beam width up to 5 is used to generate captions. 

We evaluate the performance of the proposed model using the same metrics adopted in DCASE 2021 Challenge on Task 6: ``Automated Audio Captioning", including machine translation metrics: BLEU$_{n}$ \cite{papineni2002bleu}, METEOR \cite{lavie2007meteor}, ROUGE$_{L}$ \cite{lin2004rouge} and captioning metrics: CIDEr \cite{vedantam2015cider}, SPICE \cite{anderson2016spice}, SPIDEr \cite{liu2017improved}. BLEU$_n$ measures the precision of $n$-gram inside the generated text. METEOR is a harmonic mean of precision and recall based on word-to-word matches. ROUGE$_{L}$ calculates F-measures based on the longest common sub-sequence. CIDEr considers the cosine similarity between term frequency inverse document frequency (TF-IDF) of the $n$-gram. SPICE extracts captions into scenes graphs and calculates F-score based on them. SPICE score ensures captions are semantically faithful to the audio clip, while the CIDEr score ensures captions are syntactically fluent. SPIDEr is the mean score
of CIDEr and SPICE.


\begin{table*}[ht]
\centering
\caption{Model performance on the AudioCaps dataset. Upper: performance of existing audio captioning methods (baseline). Bottom: performance of our proposed PANNs (CNN10) encoder BERT decoder model. The displayed scores are means and standard deviations over three experiments. The highest value for each metric is shown in bold.}
\begin{tabular}[\linewidth]{c c c c c c c c c c c} 
 \hline
 Model & BLEU$_{1}$ & BLEU$_{2}$ & BLEU$_{3}$ & BLEU$_{4}$ & ROUGE$_{L}$ & METEOR & CIDEr & SPICE & SPIDEr \\ 
 \hline
 TopDown-AlignedAtt \cite{kim2019audiocaps} & 61.4 & 44.6 & 31.7 & 21.9 & 45.0 & 20.3 & 59.3 & 14.4 & 36.9 \\
 CNN10-AT \cite{xu2021investigating} & 65.5 & 47.6 & 33.5 & 23.1 & 46.7 & 22.9 & 66.0 & 16.8 & 41.4 \\
 ACT\_small \cite{mei2021audio} & 64.3 & 48.3 & 35.2 & 24.9 & 46.9 & 21.8 & 66.9 & 16.0 & 41.5 \\
 ACT\_medium \cite{mei2021audio} & 65.3 & 49.5 & \textbf{36.3} & \textbf{25.9} & 47.1 & 22.2 & 66.3 & 16.3 & 41.3 \\
 ACT\_large \cite{mei2021audio} & 64.7 & 48.8 & 35.6 & 25.2 & 46.8 & 22.2 & \textbf{67.9} & 16.0 & \textbf{42.0} \\
 GPT-2 + similar captions \cite{koizumi2020audio} & 63.8 & 45.8 & 31.8 & 20.4 & 43.4 & 19.9 & 50.3 & 13.9 & 32.1 \\
 \hline
 CNN10 + BERT\_tiny & 66.0 (0.7)  & 49.1 (0.3) & 35.2 (0.3) & 24.5 (0.2) & 47.0 (0.5) & 22.4 (0.2) & 63.1 (0.9) & 16.2 (0.3) & 39.6 (0.5) \\
 CNN10 + BERT\_mini & \textbf{67.1 (0.9)} & 49.8 (0.6) & 35.8 (0.3) & 25.1 (0.1) & \textbf{48.0 (0.7)} & \textbf{23.2 (0.4)} & 66.7 (0.6) & \textbf{17.2 (0.1)} & 41.9 (0.3) \\
 CNN10 + BERT\_medium & \textbf{67.1 (0.3)} & \textbf{50.1 (0.2)} & \textbf{36.3 (0.2)} & 25.5 (0.2) & 47.9 (0.4) & 23.1 (0.4) & 65.4 (1.2) & 16.8 (0.5) & 41.1 (0.6) \\
 CNN10 + BERT\_base & 66.0 (0.4) & 48.6 (0.3) & 34.4 (0.4) & 23.7 (0.5) & 47.0 (0.2) & 22.9 (0.1) & 63.4 (1.3) & 16.5 (0.1) & 40.0 (0.6) \\
 CNN10 + RoBERTa\_base & 66.1 (0.3) & 48.6 (0.2) & 34.4 (0.2) & 23.7 (0.3) & 46.9 (0.3) & 22.3 (0.1) & 63.7 (1.6) & 16.1 (0.2) & 39.9 (0.8) \\
 \hline
\end{tabular}
\label{table:tab_results1} 
\end{table*}

\begin{table*}[ht]
\centering
\caption{The performance of the state-of-the-art system (BART + AudioSet tags) and human-generated captions (human).}
\begin{tabular}[\linewidth]{c c c c c c c c c c c} 
 \hline
 Model & BLEU$_{1}$ & BLEU$_{2}$ & BLEU$_{3}$ & BLEU$_{4}$ & ROUGE$_{L}$ & METEOR & CIDEr & SPICE & SPIDEr \\ 
 \hline
 BART + AudioSet tags \cite{gontierautomated} & 69.9 (0.5) & 52.3 (0.7) & 38.0 (0.8) & 26.6 (0.9) & 49.3 (0.4) & 24.1 (0.3) & 75.3 (0.9) & 17.6 (0.3) & 46.5 (0.6) \\
Human \cite{kim2019audiocaps} & 65.4 & 48.9 & 37.3 & 29.1 & 49.6 & 28.8 & 91.3 & 21.6 & 56.5 \\
 \hline
\end{tabular}
\label{table:tab_results2} 
\end{table*}

\begin{table*}[ht]
\centering
\caption{Performance metrics for the ablation study (randomly initialized BERT decoder). The values in the metrics where randomly initialized BERT outperforms the pre-trained BERT are in bold.}
\begin{tabular}[\linewidth]{c c c c c c c c c c c} 
 \hline
 Decoder (Random) & BLEU$_{1}$ & BLEU$_{2}$ & BLEU$_{3}$ & BLEU$_{4}$ & ROUGE$_{L}$ & METEOR & CIDEr & SPICE & SPIDEr \\ 
 \hline
 BERT\_tiny & 65.8 (0.7)  & 48.9 (0.2) & 35.0 (0.4) & 24.2 (0.4) & 47.0 (0.3) & 22.1 (0.2) & 62.2 (1.2) & \textbf{16.3 (0.1)} & 39.2 (0.7) \\
 BERT\_mini & 66.4 (0.6) & 49.2 (0.6) & 35.5 (0.5) & \textbf{25.2 (0.5)} & 47.8 (0.3) & 23.2 (0.2) & 65.3 (0.7) & 16.8 (0.2) & 41.0 (0.3) \\
 BERT\_medium & 66.7 (0.5) & 49.1 (0.6) & 35.4 (0.6) & 24.7 (0.6) & 47.5 (0.3) & \textbf{23.2 (0.3)} & 65.4 (1.0) & 16.7 (0.1) & 41.0 (0.5) \\
 BERT\_base & 64.0 (0.2) & 46.5 (0.3) & 32.4 (0.3) & 21.8 (0.1) & 45.9 (0.3) & 22.0 (0.3) & 61.0 (1.1) & 16.0 (0.0) & 38.5 (0.6) \\
 RoBERTa\_base & 64.9 (0.5) & 47.6 (0.7) & 33.5 (0.6) & 22.9 (0.4) & 46.1 (0.2) & 22.0 (0.1) & 62.0 (1.3) & \textbf{16.2 (0.4)} & 39.1 (0.6) \\
 \hline
\end{tabular}
\label{table:tab_results3} 
\end{table*}

\section{Results}
\label{sec:Results}
\subsection{Comparison with baseline methods}
We compare our proposed approach with five baseline methods, namely, the TopDown-AlignedAtt \cite{kim2019audiocaps} model, the CNN10-AT model \cite{xu2021investigating} which uses pre-trained Audio Tagging model as the encoder, the Audio Captioning Transformer (ACT) \cite{mei2021audio}, which is the first convolution-free architecture, the model in \cite{koizumi2020audio} that uses frozen GPT-2 and audio-based similar caption retrieval, and finally the current state-of-the-art model \cite{gontierautomated} on AudioCaps based on BART and AudioSet tags.

We report the performance of the first four baseline methods in the upper part of Table \ref{table:tab_results1}. It can be observed that the final method \cite{gontierautomated} cannot generalize well to other datasets. Since its performance improvement depends on the AudioSet tags used as word hinters in caption annotation stage of AudioCaps, so we separately report its performance in Table \ref{table:tab_results2} for reference.  In addition, the performance of human-generated captions described in \cite{kim2019audiocaps} is given in Table \ref{table:tab_results2}.

\subsection{Efficacy of BERT decoder}

We report the performance of our proposed model in the bottom part of Table \ref{table:tab_results1}. Experimental results demonstrate that Compact BERT achieves the best result, especially BERT\_mini and BERT\_medium. We found that although BERT and RoBERTa are more powerful pre-trained NLP models, they do not outperform Compact BERT. We empirically found that the degradation of BERT and RoBERTa is due to their large architecture, which potentially leads to over-fitting on this task. Compared with the baseline models, our models perform better on the machine translation related metrics. Specifically, BERT\_mini achieved the highest scores in BLEU$_{1}$, METEOR and CIDEr, while BERT\_medium performs the best in terms of BLEU$_{1}$, BLEU$_{2}$, BLEU$_{3}$, and ROUGE$_{L}$ metrics. This indicates that our models have a better ability in generating accurate words and fluent language descriptions than baseline models. In summary, our models show competitive performance as compared to the existing audio captioning models. 

To further show the efficacy of the BERT decoders for audio captioning, we conducted an ablation study with randomly initialized weights of the BERT decoder. Note, that there is no structural difference between a BERT decoder and a standard transformer decoder. Experimental results are reported in Table \ref{table:tab_results3}, as for all BERT architectures, the pre-trained BERT decoders outperform the randomly initialized BERT decoders on most metrics. This shows that the knowledge from the pre-trained BERT model is helpful for audio captioning.

\section{Conclusion}
\label{sec:conclusion}
We presented an encoder-decoder based audio captioning model by using pre-trained BERT models as language decoder and PANNs as the audio encoder. To bridge the language decoder and the audio encoder, the cross-attention layers are added with randomly initialized weights in the BERT decoder, while the other layers retain the pre-trained weights from BERT models. We conducted an empirical study on the utility of pre-trained BERT models with a different scale on the AudioCaps dataset. The experimental results demonstrate the efficacy of the BERT model for audio captioning. Our proposed models show competitive results as compared to the existing audio captioning methods.

\section*{Acknowledgment}

This work is partly supported by a Newton Institutional Links Award from the British Council and the Scientific and
Technological Research Council of Turkey (TUBITAK), titled ``Automated Captioning of Image and Audio for Visually and Hearing Impaired" (Grant numbers 623805725 and 120N995), a grant EP/T019751/1 from the Engineering and Physical Sciences Research Council (EPSRC), and a PhD scholarship from the University of Surrey, and a Research Scholarship from the China Scholarship Council. 

\bibliographystyle{IEEEtran}
\bibliography{refs}

\end{document}